\documentclass[a4paper,fleqn,usenatbib,useAMS]{mnras}

\usepackage[T1]{fontenc}
\usepackage{ae,aecompl}

\usepackage{graphicx}	
\usepackage{amsmath}	
\usepackage{amssymb}	
\usepackage{bm}		
\usepackage{pdflscape}	

\usepackage{times}
\usepackage{txfonts}
\usepackage{color}

\newcommand{\gw}{GW$\,$170817$\,$/$\,$GRB$\,$170817A$\,$}

\newcommand{\fracb}[2]{\left(\frac{#1}{#2}\right)}

\newcommand{\mean}[1]{\langle{#1}\rangle}

\usepackage{etoolbox}
\makeatletter
\patchcmd\@combinedblfloats{\box\@outputbox}{\unvbox\@outputbox}{}{%
 \errmessage{\noexpand\@combinedblfloats could not be patched}%
}%
\makeatother

\title[Afterglow postshock magnetic field structure]
{Constraining the magnetic field structure in collisionless relativistic shocks with a radio afterglow polarization upper limit in GW$\,$170817}
  
\author[Gill \& Granot]{
Ramandeep Gill$^{1}$\thanks{E-mail: rsgill.rg@gmail.com},
Jonathan Granot$^{1,2}$\thanks{E-mail: granot@openu.ac.il (JG)}
\\
$^{1}$Department of Natural Sciences, The Open University of Israel, P.O Box 808, Ra'anana 43537, Israel\\
$^{2}$Department of Physics, The George Washington University, Washington, DC 20052, USA\\
}

\date{Accepted XXX. Received YYY; in original form ZZZ}

\pubyear{2019}

\begin{document}
\label{firstpage}
\pagerange{\pageref{firstpage}--\pageref{lastpage}}
\maketitle

\begin{abstract}
Gamma-ray burst (GRB) afterglow arises from a relativistic shock driven into the ambient medium, 
which generates tangled magnetic fields and accelerates  relativistic electrons that radiate the 
observed synchrotron emission. In relativistic collisionless shocks the postshock magnetic field 
$\mathbfit{B}$ is produced by the two-stream and/or Weibel instabilities on plasma skin-depth 
scales $(c/\omega_p)$, and is oriented predominantly within the shock plane ($B_{\perp}$; transverse 
to the shock normal, $\hat{\mathbfit{n}}_{\rm{sh}}$), and is often approximated to be completely 
within it ($B_\parallel\equiv\hat{\mathbfit{n}}_{\rm{sh}}\,\cdot\,\mathbfit{B}=0$). Current 
2D/3D particle-in-cell simulations are limited to short timescales and box sizes 
$\lesssim10^4(c/\omega_p)\ll R/\Gamma_{\rm{sh}}$ much smaller than the shocked region's comoving 
width, and hence cannot probe the asymptotic downstream $\mathbfit{B}$ structure. We 
constrain the latter using the linear polarization upper limit, $\vert\Pi\vert<12\%$, on the radio 
afterglow of GW$\,$170817$\,$/$\,$GRB$\,$170817A. Afterglow polarization depends on the jet's angular 
structure, our viewing angle, and the $\mathbfit{B}$ structure. 
In GW$\,$170817$\,$/$\,$GRB$\,$170817A the latter can be tightly constrained since the former two 
are well-constrained by its exquisite observations. We model $\mathbfit{B}$ as an isotropic 
field in 3D that is stretched along $\hat{\mathbfit{n}}_{\rm{sh}}$ by a factor 
$\xi\equiv{}B_{\parallel}/B_{\perp}$, whose initial value $\xi_f\equiv{}B_{\parallel,f}/B_{\perp,f}$
describes the field that survives downstream on plasma scales $\ll{}R/\Gamma_{\rm{sh}}$. We calculate 
$\Pi(\xi_f)$ by integrating over the entire shocked volume for a Gaussian or power-law core-dominated 
structured jet, with a local Blandford-McKee self-similar radial profile (used for evolving $\xi$ downstream).
We find that independent of the exact jet structure, $\mathbfit{B}$ has a finite, but initially 
sub-dominant, parallel component: $0.57\lesssim\xi_f\lesssim0.89$, making it less anisotropic. While 
this motivates numerical studies of the asymptotic $\mathbfit{B}$ structure in relativistic collisionless
shocks, it may be consistent with turbulence amplified magnetic field.
\end{abstract}


\section{Introduction}


There is good evidence that synchrotron radiation is the dominant emission mechanism in most GRB afterglows  
(e.g., \citealt{Meszaros-Rees-97,Waxman-97,Sari+98}, and see \citealt{Piran-04,Kumar-Zhang-15} for a review). 
However, it depends on the rather poorly understood physics of relativistic collisionless shocks, in particular 
the microphysical processes that accelerate particles into a non-thermal energy distribution and generate 
near-equipartition tangled magnetic fields just behind the shock. These uncertainties are generally parameterized 
using the microphysical parameters, $\epsilon_e = \langle\gamma_e\rangle\rho_e c^2/e$ 
and $\epsilon_B = B^2/8\pi e$, that define the fraction of the total comoving internal energy density behind the 
afterglow shock, $e$, given to non-thermal relativistic electrons, with mean energy per unit rest-mass energy 
$\langle\gamma_e\rangle\gg1$ and comoving rest-mass density $\rho_e$, and to the magnetic field of strength $B$, 
respectively (where all of these quantities are measured in the downstream comoving rest frame, and $c$ is the 
speed of light). Detailed works comparing this synchrotron shock 
model with GRB afterglow observations find $\epsilon_e\sim10^{-1}$ and a wide range for $\epsilon_B\sim10^{-5}-10^{-1}$ 
(e.g., \citealt{Wijers-Galama-99,Panaitescu-Kumar-02}; however also see \citealt{Santana+14} who find $\epsilon_B$ values 
smaller by at least a factor of $10^{-2}$). 
These represent the emissivity-weighted mean values of the microphysical parameters assuming a uniform 
emission region, and do not account for their possible variation within the shocked region. 

The leading theoretical explanation for magnetic field generation at the collisionless relativistic forward shock posits 
that when the upstream plasma into which the shock is expanding is weakly magnetized or unmagnetized (with a magnetization 
parameter $\sigma=B^2/4\pi\rho c^2\lesssim10^{-3}$), magnetic fields are 
produced by the relativistic two-stream and/or Weibel (filamentation) instabilities 
\citep{Weibel-59,Gruzinov-Waxman-99,Medvedev-Loeb-99,Bret09,Keshet+09,Nakar+11}. The shock-accelerated supra-thermal particles 
that escape into the upstream plasma and propagate ahead of the shock excite micro-instabilities in a spatially extended region 
called the \textit{precursor}. These micro-instabilities (e.g. the Weibel-filamentation) generate a magnetic barrier at the proton 
skin-depth scales of $c/\omega_p=2.3\times10^7n_0^{-1/2}\,$cm, where $c$ is the speed of light, $\omega_p$ 
is the proton plasma frequency, and $n_0 = n/(1~{\rm cm}^{-3})$ is the upstream particle number density, which grows to 
near-equipartition with $\epsilon_B\sim10^{-2}-10^{-1}$ and isotropizes the incoming plasma (\citealt{Moiseev-Sagdeev-63}; see, e.g. 
\citealt{Sironi+15} for a review). 
The generated field is randomly oriented but lies predominantly in the plane transverse to the direction of shock propagation. Since the size of the 
emission region (below the cooling frequency) in the comoving frame $\Delta_{\rm sh}'$ is much larger, with 
$(c/\omega_p)\ll\Delta_{\rm sh}'\sim R/\Gamma_{\rm sh} = 10^{14}R_{15}\Gamma_{\rm sh,1}^{-1}\,$cm where $R$ and $\Gamma_{\rm sh}$ are, 
respectively, the radial distance and Lorentz factor (LF) of the shock, 
the field must be able to survive much deeper downstream of the shock \citep{Piran-05} without completely decaying due to 
particle phase-space mixing \citep{Gruzinov-01}. 

Numerical simulations and analytic works show that the coherence scale of the magnetic field grows beyond the skin-depth scales via the 
formation of current filaments \citep{Silva+03,Frederiksen+04,Medvedev+05}. However, these may be subject to pressure-driven instabilities, 
e.g. the kink instability, which would destroy the filamentary structure, thermalize the particles, and cause the field to decay in the 
shocked region \citep{Milosavljevic-Nakar-06}. 2D $e^\pm$-pair plasma PIC simulations \citep{Chang+08,Spitkovsky-08a} and 2.5D electron-ion 
PIC simulations \citep{Spitkovsky-08b} have also found that the filaments break up into clumps surrounded by a highly isotropic plasma and 
the magnetic field rapidly decays after $\sim$few$\,\times\,100\,(c/\omega_{p,e})$, with $\omega_{p,e}$ being the electron plasma frequency, 
and $\sim20(c/\omega_p)$, respectively. Many of these results are derived from the short-term evolution of the shock structure and the downstream 
magnetic field. Long-term PIC simulations \citep[e.g.,][]{Keshet+09} instead find that at times $t>10^3\omega_{p,e}^{-1}$ the properties of 
the shock and current filaments in the precursor region are gradually modified by shock-accelerated energetic particles. This causes a gradual 
increase in the level of magnetization and the magnetic field coherence scale in the upstream. Consequently, as the magnetic field advects 
into the downstream, the decay rate of the postshock magnetic field slows down, its coherence length grows, and $\epsilon_B$ approaches $\sim10^{-2}$ on 
length scales up to $\sim10^3c/\omega_{p,e}$ downstream of the shock transition.

Synchrotron radiation is partially linearly polarized, and therefore, measurement of linear polarization of GRB afterglows is an 
invaluable tool to study the asymptotic structure of the postshock magnetic field. However, the emergent polarization depends not only on the 
magnetic field structure but also on the structure of the jet and the observer's line-of-sight (LOS), leading to considerable degeneracy for an 
off-axis ($\theta_{\rm obs}>0$) observer. To break the degeneracy between the magnetic field structure, the jet structure and $\theta_{\rm obs}$, it is important to first independently model 
the latter two using the afterglow lightcurve and image on the plane of the sky. In this work, we use the exquisite broadband afterglow 
observations of \gw and the semi-analytic model fits (from afterglow data up to $t_{\rm obs}\sim600\,$ days) obtained by 
\citet{Gill-Granot-18b,Gill+19} for axi-symmetric core-dominated jet angular 
structures. These models for the jet's angular structure and $\theta_{\rm obs}$ are then used to predict the degree of linear polarization 
for different tangled postshock magnetic fields, which are naturally symmetric with respect to $\hat{\mathbfit{n}}_{\rm sh}$. Finally, our model 
predictions are compared to the polarization upper limit for \gw to constrain the postshock magnetic field structure.

The rest of the paper is organized as follows. In \S\ref{sec:degeneracy}, we first present a general discussion of how linear 
polarization is produced in axisymmetric flows from different magnetic field configurations and jet structures. Then, we briefly 
discuss how afterglow observations of \gw were used to constrain the jet structure and $\theta_{\rm obs}$. In \S\ref{sec:jet-model} 
we describe the jet structure and dynamics used in this work, which features an angular structure given by two semi-analytic models
of core-dominated structured jets with local spherical dynamics \citep[see][for more details]{Gill-Granot-18b} superimposed  
with a \citet{Blandford-McKee-76} radial profile for the postshock flow. In \S\ref{sec:B-field}, in order to calculate the linear 
polarization, we model the postshock magnetic field $\mathbfit{B}$ and parameterize its degree of anisotropy through 
$\xi\equiv{}B_{\parallel}/B_{\perp}$, whose initial value $\xi_f\equiv{}B_{\parallel,f}/B_{\perp,f}$ 
describes the field that survives downstream on plasma scales $\ll{}R/\Gamma_{\rm{sh}}$. Since each fluid element is stretched more 
along the shock normal direction $\hat{\mathbfit{n}}_{\rm sh}$ than in the two perpendicular directions as it flows downstream, 
$\xi$ grows with the distance behind the shock. In \S\ref{sec:lin-pol} we obtain strong constraints on the shock-generated field 
anisotropy just behind the shock (namely $0.57\lesssim\xi_f\lesssim0.89$) by comparing the predicted degree of polarization to the 
radio upper limit. Finally, in \S\ref{sec:discussion} we discuss the important implications that this result may have for the magnetic 
fields generated in relativistic collisionless shocks. Our results strongly suggest that the shock-generated field must have 
a component parallel to the shock normal and it cannot be only in the plane transverse to it, as suggested in earlier works 
\cite[e.g.,][]{Medvedev-Loeb-99}. We find the postshock field to be less anisotropic, which may be consistent with a turbulence 
amplified magnetic field.

Throughout this work, the notation $Q_x$ denotes the value of the quantity $Q$ in units of $10^x$ times its (cgs) units.

\section{Linear polarization of GRB afterglows: magnetic field and jet structures}
\label{sec:degeneracy}
Here we summarize the different ways in which net linear polarization is obtained in GRB afterglows. We point out the degeneracy 
between different magnetic field configurations that can potentially lead to similar levels of polarization. This is further 
complicated by the degeneracy between the magnetic field configuration, jet structure and viewing angle $\theta_{\rm obs}$, where 
off-axis viewing ($\theta_{\rm obs}>0$) breaks the symmetry of the image for an axisymmetric flow, leading to net polarization.

Since GRBs are cosmological sources and their images 
on the plane of the sky are generally unresolved, any measurement of linear polarization effectively averages the local polarization 
over the entire image. As a result, linear polarization from shock-generated fields, which are on average symmetric around the local 
shock normal direction, $\hat{\mathbfit{n}}_{\rm{sh}}$ (as it is the only relevant preferred direction locally), 
would average out for a spherical flow and produce no net polarization (as there would be no global preferred direction).\footnote{Magnetic fields that are randomly oriented in 3D space 
would yield zero polarization even locally as there is no preferred direction for the polarization vector.} Therefore, in order to 
detect any residual polarization, the symmetry of the image must be broken. For a spherical flow this can occur if different weight 
is given to different parts of the image, e.g. through microlensing \citep{Loeb-Perna-98}, radio scintillations \citep{Medvedev-Loeb-99}, 
clumps in the external medium \citep{Granot-Konigl-03}, or angular inhomogeneities in the jet \citep[``patchy shell'',][]{Kumar-Piran-00,Ioka-Nakamura-01,Granot-Konigl-03,Nakar-Oren-04,Yamazaki+04}. 
However, these are expected to occur only in some fraction of GRB afterglow, and would be accompanied by temporal variability in the 
afterglow lightcurve that was not observed in GRB$\,$170817A. 

A more robust and therefore also more popular way of breaking the symmetry of the afterglow image is through an axisymmetric flow (i.e. a jet) 
observed off-axis (from $\theta_{\rm obs}>0$, where $\theta_{\rm obs}$ is measured from the jet's symmetry axis). This can occur in two ways: 
(i) if the emission arises from a relativistic (with LF $\Gamma\gg1$) uniform jet with sharp edges at an initial half-opening angle $\theta_0$ 
(i.e. a ``top-hat'' jet), an off-axis observer within the jet's aperture ($0<\theta_{\rm{obs}}<\theta_0$) sees the edge of the jet near the 
time of the jet break in the lightcurve, when the flow decelerates so that the beaming cone of the emission widens to 
$1/\Gamma\gtrsim(\theta_0-\theta_{\rm obs})$ \citep{Ghisellini-Lazzati-99,Sari-99}. This results in incomplete cancellation of polarization which 
yields finite residual polarization when averaged over the GRB image; (ii) If the flow is structured and its properties, e.g. the energy per unit 
solid angle and/or Lorentz factor, vary with polar angle $\theta$ from the jet symmetry axis \citep[i.e. a ``structured'' jet, e.g.][]{Kumar-Granot-03}, 
the gradient in the polarized intensity in the observed region again leads to incomplete cancellation of polarization \citep{Rossi+04,Gill-Granot-18b}.

Alternatively, net polarization is obtained if the shocked region is permeated by an ordered magnetic field component in addition to the 
shock-generated random field \citep{Granot-Konigl-03}, or if the emission region consists of coherent magnetic field patches \citep{Gruzinov-Waxman-99} 
of angular scale $\theta_B<1/\Gamma$, such that the image (i.e. the observed region of angle $\sim1/\Gamma$ around our line of sight) contains 
$N\sim(\Gamma\theta_B)^2$ patches. In these two cases, the symmetry is broken by the ordered field component, for which the local maximum polarization is 
$0.5\leq[\Pi_{\max}=(\alpha+1)/(\alpha+5/3)]\lesssim0.75$ where $\alpha=-d\log F_\nu/d\log\nu$ is the spectral index. In the $N$-patches model, 
since the emission arises from $N$ intrinsically coherent but mutually incoherent patches, the net polarization is reduced to $\Pi\sim\Pi_{\max}/\sqrt{N}$ 
due to partial cancellation (the $\sqrt{N}$ suppression factor arising from random walk in the Stokes parameters ($U,\,Q$) plane).

Measurement of $\Pi\sim1\%-3\%$ in the optical and NIR afterglow of several GRBs (e.g., \citealt{Covino+99,Wijers+99}, 
also see \citealt{Covino-Gotz-16} for a review) confirmed the synchrotron origin of the emission. Such low levels of polarization 
readily ruled out an ordered magnetic field with coherence angular scale $\theta_B\gtrsim1/\Gamma$. Instead, they suggested either ordered fields 
with much smaller coherence length scales, as in the $N$-patches scenario, or shock-generated tangled fields. In the latter case, 
however, the magnetic field must have some level of anisotropy if the 
flow is uniform or the GRB image must be inhomogeneous owing to the fact that the observer is off-axis and the jet either has a 
sharp edge or a core-dominated angular structure. \citet{Granot-Konigl-03} parameterized the postshock tangled magnetic field 
anisotropy using $b \equiv 2\langle B_\parallel^2\rangle/\langle B_\perp^2\rangle$, the ratio of the radially averaged energy densities 
of the two field components. Note that $b=1$ for a field that is isotropic in 3D and gives zero local polarization, so the closer 
$b$ is to one the lower the local and global degrees of polarizations. Using the observed levels of afterglow polarization $\Pi\sim1\%-3\%$ 
around several hours to a few days, which were relatively close to the jet break time around which the polarization is expected to peak, 
and assuming emission from an infinitely thin relativistic spherical shell, \citet{Granot-Konigl-03} constrained $b$ to be approximately 
within the range $0.5\lesssim b\lesssim2$. This involved a statistical argument about the distribution of $\theta_{\rm obs}$ between the 
different GRBs within the modest sized sample with afterglow polarization measuremants that was available in 2003. Nevertheless, it 
already suggested that the postshock magnetic field must be at least mildly anisotropic. 

Still, considerable degeneracy remains between the scenarios mentioned above and one way to break that is by the observation of the temporal evolution 
of the polarization position angle (PA), $\theta_p$. In the $N$-patches case, $\theta_p$ is expected to vary randomly and $\Pi$ fluctuates 
and gradually decreases, both over timescales $\Delta t\sim t$, as the visible region grows and encompasses more patches. Interestingly, 
both of these tell-tale signatures were recently observed by ALMA at $97\;$GHz in the reverse-shock emission (from the shocked original 
GRB ejecta) of GRB$\,$190114C between 2.2 and 5.2 hours after the GRB, implying $\theta_B\sim10^{-3}\;$rad \citep{Laskar+19}. On the other 
hand, for a mildly anisotropic shock generated magnetic field, $\theta_p$ flips by $90^\circ$ as $\Pi$ momentarily vanishes, around the time 
of the jet-break for a top-hat jet viewed off-axis \citep{Ghisellini-Lazzati-99,Sari-99,Granot-Konigl-03}; for a structured jet viewed off-axis 
$\theta_p$ remains unchanged \citep{Rossi+04}.

Since $\theta_{\rm obs}$, the jet angular structure and 
the degree of anisotropy of the shock-generated magnetic field, all affect the observed polarization, 
it becomes important to model the angular structure of the flow, and constrain $\theta_{\rm{obs}}$. Almost all GRBs, whether of the long-soft or short-hard classes, have been observed 
at cosmological distances, which guarantees that the observer's LOS lies within the beaming cone of the compact core of half-opening angle $\theta_c$, such that $\theta_{\rm obs}-\theta_c\lesssim\rm{few}\times\Gamma^{-1}$ for core-dominated jets where the emission sharply drops at $\theta>\theta_c$. This makes it challenging to draw any useful 
inferences on the jet structure and on our viewing angle $\theta_{\rm obs}$. 

The afterglow of the short-hard GRB 170817A \citep{Abbott+17-GW170817-GRB170817A}, associated to the first-ever detection of a binary 
neutron star merger gravitational wave source GW$\,$170817 \citep{Abbott+17-GW170817-Ligo-Detection}, has provided a golden opportunity 
to study the structure of outflows that power short-hard GRBs. The broadband afterglow, detected after $8.9\,$days in X-rays \citep{Troja+17} 
and $16.4\,$days in the radio \citep{Hallinan+17}, showed an unusually long-lasting flux rise, as 
$F_\nu(t_{\rm{obs}})\propto\nu^{-0.6}t_{\rm{obs}}^{0.8}$, up to the peak at $t_{\rm{obs,pk}}\sim 150\,$days post merger 
\citep[e.g.][]{Lyman+18,Margutti+18,Mooley+18a}, followed by a sharp decay as $F_\nu\propto{}t_{\rm{obs}}^{a}$ where $a\simeq-2$ 
\citep{Mooley+18b,vanEerten+18,Hajela+2019,Lamb+19}. Several numerical and semi-analytic works modeled the afterglow as arising 
from a successful (i.e. bore its way through the merger dynamical ejecta rather than being chocked within it) off-axis 
core-dominated structured jet, whose angular profile is consistent with either a (quasi-) Gaussian or a narrow core with sharp power-law wings 
\citep{Alexander+18,Lamb-Kobayashi-18,Lazzati+18,Nynka+18,Troja+17,DAvanzo+18,Gill-Granot-18b,Lyman+18,Margutti+18,Resmi+18,Troja+18,Hajela+2019,Lamb+19}. Later radio VLBI observations measured 
an apparent superluminal motion of the afterglow flux centroid with $\varv_{\rm app}\approx4c$ \citep{Mooley+18b}, 
and constrained the angular size of the afterglow image on the plane of the sky to $\lesssim2\,$mas \citep{Ghirlanda+18,Mooley+18b}. The 
apparent superluminal motion firmly established the fact that the outflow had a narrow relativistic compact core surrounded by low energy wings, 
which is consistent with the upper limit on its angular size. 

Using VLA radio observations at $2.8\;$GHz an upper limit of $\vert\Pi\vert<12\%$ ($99\%$ confidence) was obtained on the 
afterglow linear polarization at $t_{\rm obs}\approx244\,$days \citep{Corsi+18}. Comparison with detailed predictions for 
the linear polarization from semi-analytic models of core-dominated structured jets that explained the afterglow lightcurve 
and image size of GW$\,$170817/GRB$\,$170817A \citep{Gill-Granot-18b}, revealed that $0.7\lesssim b\lesssim1.5$, suggesting 
that the postshock magnetic field must be at most mildly anisotropic with a finite magnetic field component in the 
direction of the shock normal that is comparable to the components in the two perpendicular directions \citep{Corsi+18,Gill+18}. 
Having constrained the jet structure and viewing angle from the afterglow lightcurve and image properties of \gw, this new tighter 
constraint on $b$ was free of any model degeneracy due to the jet structure and provided a robust estimate of the radially averaged 
magnetic field anisotropy. This severely constrained models of Weibel-instability-generated fields at collisionless shocks, where 
the field lies completely in the plane transverse to the shock normal. However, the calculation in \citet{Gill-Granot-18b} did not 
account for the radial structure of the flow and it effects on the magnetic field structure, since they adopted the thin-shell 
approximation. Therefore, the upper limit on linear polarization could only be used to constrain the radially averaged degree of anisotropy.

The \gw\ broadband afterglow spectrum was explained by a single power law segment (PLS) of synchrotron emission -- PLS G 
\citep{Granot-Sari-02} with spectral index $\alpha=(p-1)/2$ (with $F_\nu\propto\nu^{-\alpha}$ and power-law relativistic 
electron distribution $N(\gamma)\propto\gamma^{-p}$). In PLS G the emission is from slow-cooling electrons and hence it 
arises from the entire shocked volume behind the forward shock. Therefore, in this work we obtain the afterglow linear polarization 
by integrating over the entire emitting volume behind the forward shock.

\section{Core-dominated angular structured flow with Blandford-McKee radial profile}
\label{sec:jet-model}

\begin{figure}
    \centering
    \includegraphics[width=0.35\textwidth]{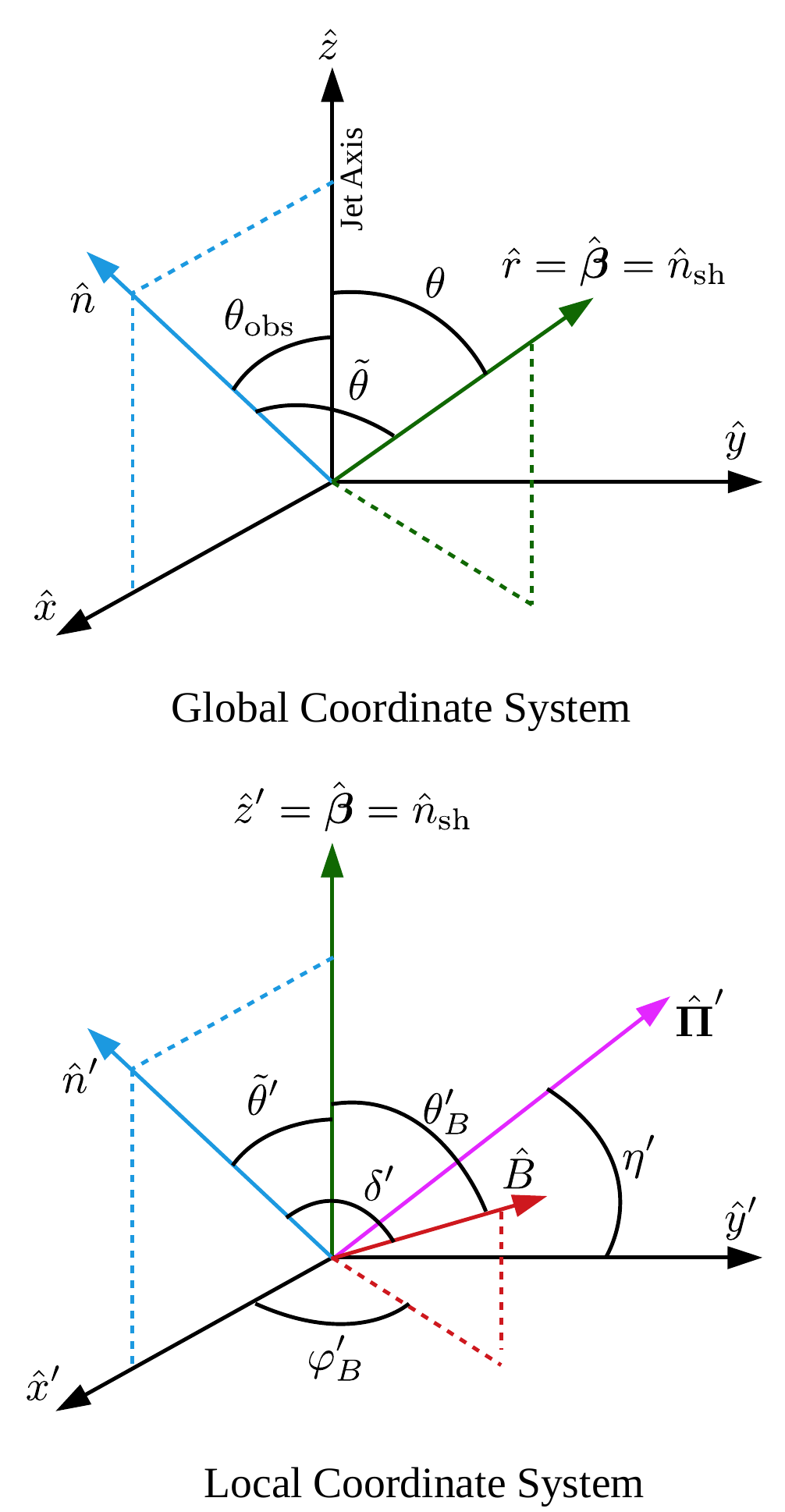}
    \caption{The global (lab-frame) and local (comoving-frame) coordinate systems, where the former is 
    used to describe the structure of the jet, and the latter is used to describe the local magnetic field 
    and is used in the calculation of linear polarization.}
    \label{fig:coord-sys}
\end{figure}

We consider a core-dominated structured jet \citep[e.g.,][]{Meszaros+98} with energy per unit solid angle,
$\mathcal{E}(\theta)\equiv dE(\theta)/d\Omega$, and initial bulk $\Gamma_0(\theta)$ of the shocked material just 
behind the shock front, both declining with the polar angle $\theta$ from the 
jet symmetry axis (see Fig.~\ref{fig:coord-sys}). Here we consider two distinct angular profiles: (i) A \textit{Gaussian jet} 
(GJ) for which both $\mathcal{E}(\theta)$ and the initial kinetic energy per unit rest mass, $\Gamma_0(\theta)-1$, 
have a Gaussian profile \citep{Zhang-Meszaros-02,Kumar-Granot-03,Rossi+04}
with a floor at $\theta=\theta_*$ corresponding to  $\beta_0=\beta_{\rm min} = 0.01$,
\begin{align}
({\rm GJ})\quad\quad\frac{\mathcal{E}(\theta)}{\mathcal{E}_c} = \frac{\Gamma_0(\theta)-1}{\Gamma_c-1} 
= \exp\left[-\frac{1}{2}\fracb{\min\{\theta,\,\theta_*\}}{\theta_c}^2\right]~,
\end{align}
where $\mathcal{E}_c$ and $\Gamma_c$ represent the core values, 
and (ii) a \textit{Power-law jet} (PLJ) for which both $\mathcal{E}(\theta)$ and $\Gamma_0(\theta)-1$ 
decline as a power law outside of the core,
\begin{align}
({\rm PLJ})\quad\ \ \frac{\mathcal{E}(\theta)}{\mathcal{E}_c}=\Theta^{-a},\quad\ \  \frac{\Gamma_0(\theta)-1}{\Gamma_c-1}=\Theta^{-b},
\quad\ \  \Theta = \sqrt{1+\fracb{\theta}{\theta_c}^2}~,
\end{align}
with their respective power law indices, $a$ and $b$.

For the dynamics, we neglect sideways expansion (or any temporal evolution of $\mathcal{E}(\theta)$), and assume that 
each point on the shock front expands locally only radially as 
part of a spherical flow with the local value of $\mathcal{E}(\theta)$.
The rest mass density of the circumburst medium (CBM) is assumed to vary as a power-law with the radial distance $r$ from the central source,    
$\rho_k(r) = m_pn(r) = A_kr^{-k}$, where $k=0$  for a 
uniform density interstellar medium (ISM) and $k=2$ for a steady wind medium, $n(r)$ is the CBM particle number density, and $m_p$ is 
the proton mass. When the forward shock reaches a radius $r$, the corresponding spherical flow would have swept up a mass $m(r) = [4\pi A/(3-k)]r^{3-k}$, and since $m(r)$ increase with $r$ this decelerates the shock.  
A wind-like environment ($k=2$) is only relevant for GRBs of the long-soft class, where quasi-steady mass loss due to stellar winds from the progenitor star may be expected.
In the case of BNS mergers, which is relevant for this work, a uniform ISM-like CBM ($k=0$) is expected.

At an early stage, the shell is assumed to coast at a constant proper velocity 
$u_0(\theta) = \Gamma_0(\theta)\beta_0(\theta)$
until the deceleration radius, 
which is expressed as
\begin{equation}
R_d(\theta) = \left[\frac{(3-k)E_{\rm k,iso}(\theta)}{4\pi Ac^2u_0^2(\theta)}\right]^{1/(3-k)}~.
\end{equation}
At the deceleration radius most of the isotropic equivalent energy of the blast wave, 
$E_{\rm k, iso}(\theta) = 4\pi\mathcal{E}(\theta)$, is used up to accelerate the 
swept up mass to $u\approx u_0(\theta)$, and also to heat it up to a similar thermal proper 
velocity, so that $m[r_d(\theta)]u_0(\theta)c^2=E_{\rm k, iso}(\theta)$. Beyond this radius, 
the shell starts to decelerate as it continues to sweep up more mass and its dynamical evolution 
becomes self-similar, such that $u(\theta) \propto r^{-(3-k)/2}$, which is valid not only in 
the relativistic phase but also in the Newtonian Sedov-Taylor phase. During this deceleration 
phase, the dynamically averaged bulk LF of the postshock material can be expressed as \citep{Panaitescu-Kumar-00,Gill-Granot-18b}
\begin{equation}\label{eq:Gamma_f}
\tilde\Gamma(\theta,\tilde{r}) = \frac{\Gamma_0(\theta)+1}{2\,\tilde{r}^{\,3-k}}
\left[\sqrt{1+\frac{4\Gamma_0(\theta)}{\Gamma_0(\theta)+1}\tilde{r}^{\,3-k}+
\left(\frac{2\tilde{r}^{\,3-k}}{\Gamma_0(\theta)+1}\right)^2}-1\right]~,
\end{equation}
where $\tilde{r}=R/R_d(\theta)$. This estimate is particularly useful when the radiating electrons in 
the shocked material are fast cooling so that the lab-frame size of the emission region is much smaller than the 
width of the shocked region, $\Delta_\gamma \ll \Delta_{\rm sh}\sim R/\Gamma_{\rm sh}^2$. 
In this case, the emission region is generally approximated as being infinitely thin with bulk LF of 
the shell given by Eq.~(\ref{eq:Gamma_f}). From the \citet[][BM76 hereafter]{Blandford-McKee-76} solution, 
the LF of the material just behind the shock is given by $\Gamma_f=[(17-4k)/4(3-k)]^{1/2}\tilde\Gamma$. 
Here and what follows, the subscript $f$ indicates the magnitude of any quantity just behind the shock front. 
So that the expression for $\Gamma_f$ remains valid in the non-relativistic regime as well, we use more generally 
$u_f=\Gamma_f\beta_f=(\Gamma_f^2-1)^{1/2}=[(17-4k)/4(3-k)]^{1/2}\tilde u$, 
where $\tilde u = \tilde\Gamma\,\tilde\beta$. The LF of the shock front is then given by (BM76)
\begin{equation}
\Gamma_{\rm sh}^2 = \frac{(\Gamma_f+1)[\hat\gamma_f(\Gamma_f-1)+1]^2}
{\hat\gamma_f(2-\hat\gamma_f)(\Gamma_f-1)+2}\approx2\Gamma_f^2\quad{\rm for}\quad\Gamma_f\gg1\,,
\end{equation}
where the adiabatic index $\hat\gamma_f = 4/3~(5/3)$ for a relativistic (Newtonian) shock.

\begin{figure}
    \centering
    \includegraphics[width=0.477\textwidth]{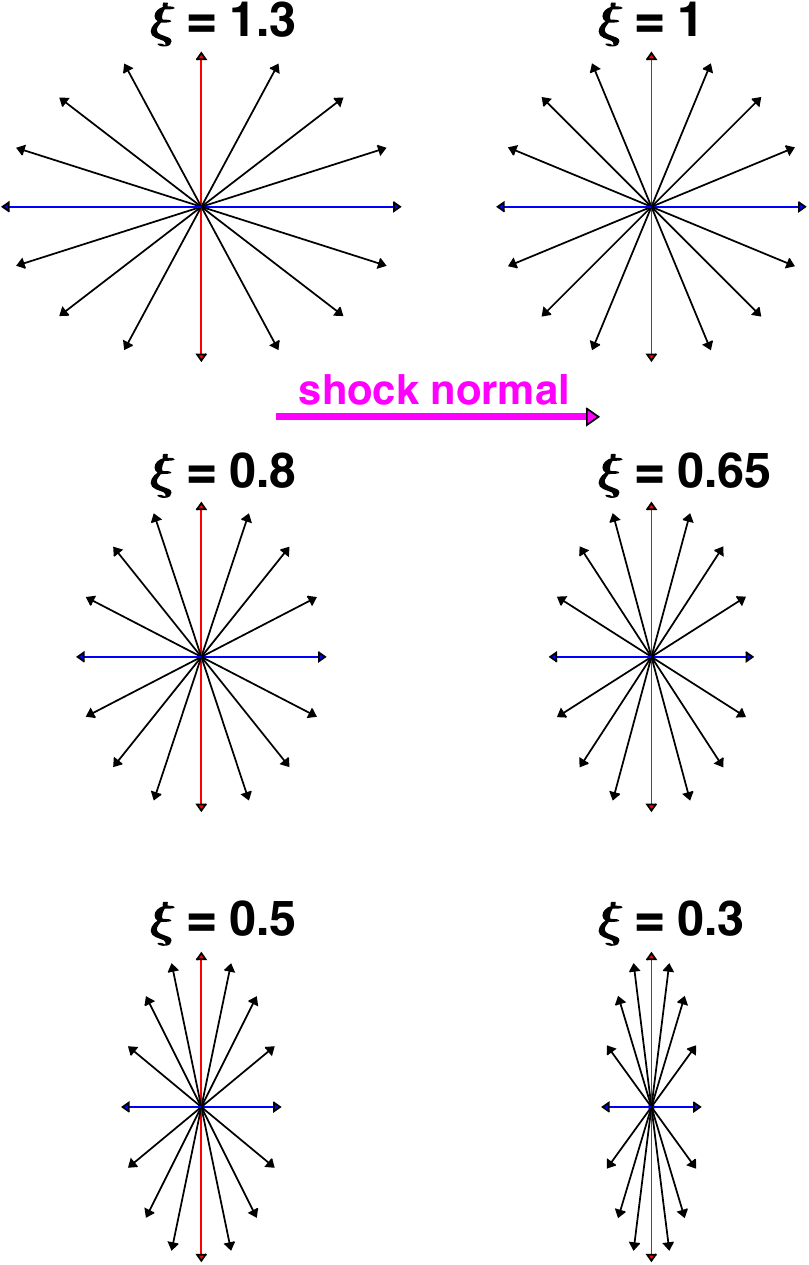}\quad
    \includegraphics[width=0.477\textwidth]{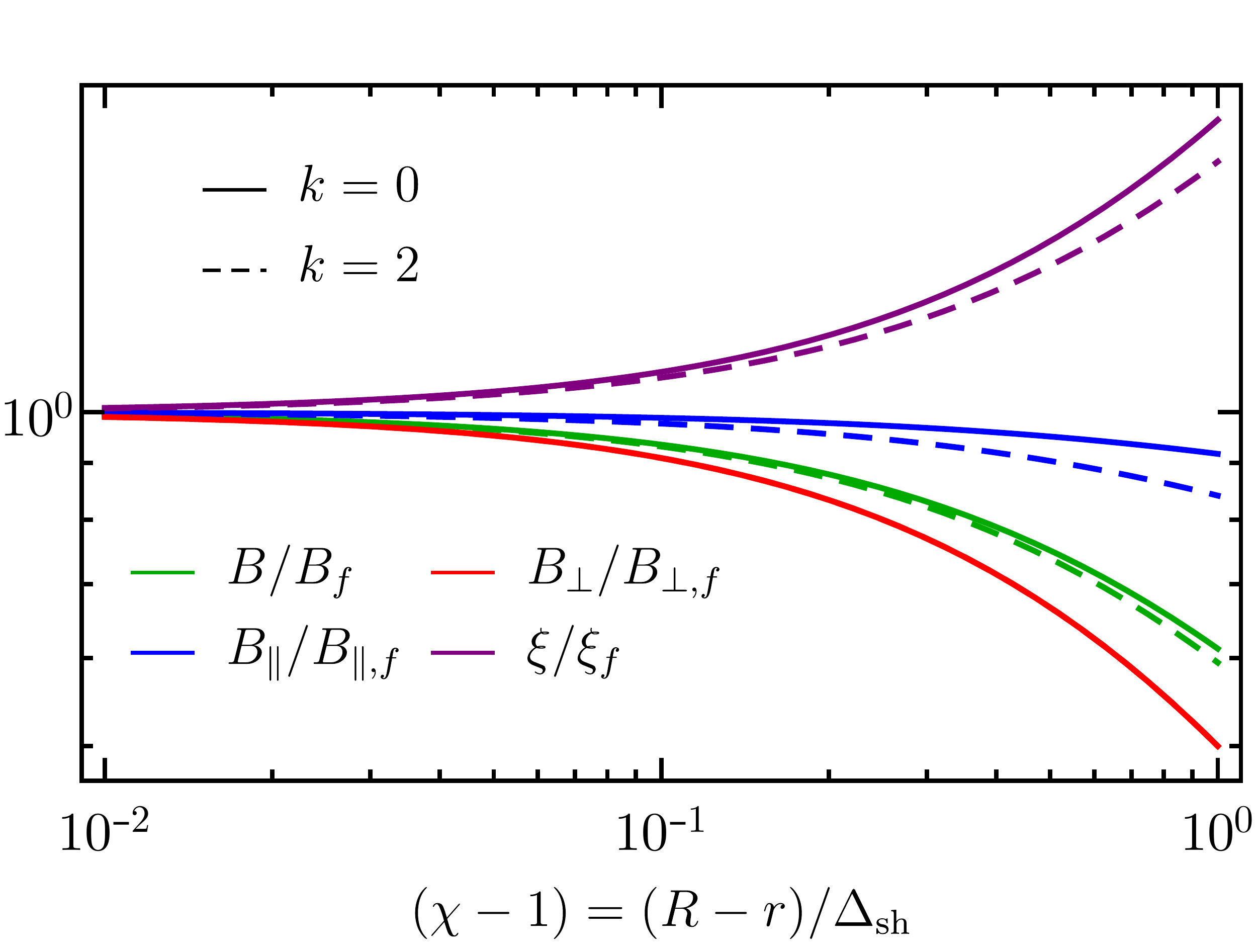}
    \caption{\textbf{Top}: Schematic of postshock magnetic field geometry for different values 
    of the anisotropy parameter $\xi = B_\parallel/B_\perp = \xi_f\,\chi^{(7-2k)/(8-2k)}$. 
    \textbf{Bottom}: Magnetic field radial profile shown as a function of the radial 
    distance behind the shock, $(R-r)$, normalized by the lab-frame width of the shocked region, 
    $\Delta_{\rm sh}\equiv R/[2(4-k)\Gamma_{\rm sh}^2]$. Here $r$ is the 
    radial coordinate and $R$ is the radial distance of the shock front. All quantities are normalized 
    by their value immediately behind the shock.}
    \label{fig:B-profile}
\end{figure}

When the postshock electrons that dominate the emission in the observed frequency range are slow cooling (i.e. cool on a timescale larger than the dynamical time), then the emission is no longer limited to a very thin layer behind 
the forward shock. Instead, it arises from a larger volume of the shocked region, containing most of the energy and swept-up mass. 
In this case, it becomes 
important to know the properties of the emitting material downstream of the shock. It was shown by BM76 that 
at $R>R_d$ the dynamics of a spherical blast wave become self-similar, such that the radial profile of 
the postshock fluid can be described using a similarity variable
\begin{equation}
    \chi = 1+2(4-k)\Gamma_{\rm sh}^2(1-y)~,
\end{equation}
where $y = r/R$ is the fractional radius 
and $r$ is the radial distance from the central source. 
For an adiabatic blast wave with impulsive energy injection, the proper mass and energy densities, 
and the proper velocity of the downstream shocked material evolve with $\chi$, such that 
\citep[e.g. BM76;][]{Granot-Sari-02,DeColle+12a}
\begin{eqnarray}\label{eq:rho-chi}
    \rho &=& 2^{3/2}\rho_k(r)\Gamma_{\rm sh}\chi^{-(10-3k)/[2(4-k)]} \\
    e &=& 2\rho_k(r)c^2\Gamma_{\rm sh}^2\chi^{-(17-4k)/[3(4-k)]} \\
    u &=& \Gamma\beta = u_f\chi^{-1/2}~.\label{eq:u-chi}
\end{eqnarray}
The downstream electron proper number density is equal to that of the protons, $n_e = n = \rho/m_p$.
The radial dependence of $u$ in Eq.~(\ref{eq:u-chi}), which is derived from the radial dependence of $\Gamma$ for the BM76 solution, 
strictly holds only in the ultrarelativistic regime, with $u\approx\Gamma\gg1$. Here we assume that the same dependence also holds in 
the transrelativistic regime as well. As shown below, this approximation has very little affect on the final result since the flux at 
the relevant times is dominated by the relativistic core where the BM76 solution still holds.

\section{Postshock Magnetic Field Structure}
\label{sec:B-field}
Immediately behind the forward shock, the energy density imparted to the magnetic field 
can be parameterized in the standard way, where a fraction $\epsilon_B$ of the total internal energy 
density $e$ goes to the magnetic field. Then, downstream of the shock the strength of the 
comoving magnetic field would evolve with the similarity variable, such that 
\begin{equation}\label{eq:B-chi}
    \frac{B}{B_f} = \left(\frac{\epsilon_B}{\epsilon_{B,f}}\frac{e}{e_f}\right)^{1/2} 
    = \fracb{\epsilon_B}{\epsilon_{B,f}}^{1/2}\chi^{-(17-4k)/(24-6k)}
\end{equation}
at a given lab frame time $t$ or shock radius $R$. For convenience it is typically assumed that $\epsilon_B = \epsilon_{B,f}$, 
but nothing guarantees that this holds in the entire downstream region. Furthermore, apart from the above scaling it is less clear how the structure 
of the magnetic field evolves downstream of the shock.


Further insight can be gained by making the assumption, without loss of 
generality, that the magnetic field just behind the shock has a component parallel ($B_{\parallel,f}$) 
and transverse ($B_{\perp,f}$) to the unit vector in the direction of the shock normal $\hat{\mathbfit{n}}_{\rm sh}$, 
which is identified here with the radial velocity unit vector 
$\hat{\mathbfit{v}}=\hat{\mathbfit{r}}=\hat{\mathbfit{n}}_{\rm sh}$. 
Here we follow the parameterization of the postshock magnetic field in \citet{Granot+99} who considered 
a spherical flow, for which the parallel direction is the radial direction. Under the ``frozen field'' approximation, 
and for a radially expanding flow, the two components of the magnetic field would evolve with $\chi$, so that 
\begin{equation}\label{eq:B-perp-par-chi}
 B_\parallel(\chi) = B_{\parallel,f}\,\chi^{-1/(8-2k)}\quad{\rm and}\quad B_\perp(\chi) = B_{\perp,f}\,\chi^{-1}~.
\end{equation}
We provide further details on the evolution of the magnetic field downstream of the shock in appendix~(\ref{sec:appendix}).

In general, the magnetic field can also have an angular distribution. \citet{Sari-99} provided a general description 
of the magnetic field anisotropy, allowing for a dependence of the (comoving) magnetic field strength of the (comoving) angle 
$\theta'_B = \arccos(\hat{\mathbfit{B}}\cdot\hat{\mathbfit{n}}_{\rm sh})$
from the shock normal $\hat{\mathbfit{n}}_{\rm sh}$, $B=B(\theta'_B)$, 
as well as a probability per unit solid angle $f(\theta'_B)$ for the magnetic field to be inclined by an angle $\theta'_B$ 
(see Fig.~\ref{fig:coord-sys} for the local coordinate system used to describe the magnetic field geometry and to calculate 
linear polarization). 
He further suggested a useful realization, which we adopt here, that takes an isotropic field and multiplies the component along 
$\hat{\mathbfit{n}}_{\rm sh}$ by a factor $\xi$, i.e. 
$B_\parallel=\mathbfit{B}\cdot\hat{\mathbfit{n}}_{\rm sh}\to\xi B_\parallel$, which leads to
\begin{equation}\label{eq:B_f}
 B(\theta_B') \propto (\xi^2\sin^2\theta_B'+\cos^2\theta_B')^{-1/2}\quad {\rm and}\quad f(\theta_B')\propto B^3(\theta_B')~.
\end{equation}
Here the two magnetic field components are described as $B_\parallel = \xi\bar B_\parallel \propto \xi\cos\theta_B'$ and 
$B_\perp = \bar B_\perp \propto \sin\theta_B'$, where the $B_\parallel$ component is scaled with respect to the unscaled component 
($\bar B_\parallel$) by $\xi$, the parameter that controls the degree of anisotropy. The $B_\perp$ component is left unscaled. 
This implies that when $\xi\gg1$ ($\xi\ll1$), $B_\parallel\gg B_\perp$ ($B_\parallel\ll B_\perp$) and $\xi=1$ produces a 
completely isotropic field.

Since the magnitude of the two field components evolves downstream of the shock, the total field anisotropy must also 
depend on the similarity variable $\chi$. The scaling of $\xi$ from an initial value of $\xi_f$ just behind the shock can 
be derived by taking the ratio of the two field components given in Eq.~(\ref{eq:B-perp-par-chi}), which yields
\begin{equation}\label{eq:xi}
    \xi = \frac{B_\parallel(\chi)}{B_\perp(\chi)} = \xi_f\chi^\frac{7-2k}{8-2k}~,\quad\quad
\xi_f=\frac{B_{\parallel,f}}{B_{\perp,f}}~.
\end{equation}
The comoving magnetic field strength depends both on its inclination angle $\theta'_B$ and on its 
radial distance behind the shock (through $\chi$) in addition to $\xi_f$, 
$B = B(\xi_f,\chi,\theta_B')$, as does its angular probability distribution according to Eq.~(\ref{eq:B_f}), 
$f=f(\xi_f,\chi,\theta_B')\propto B^3(\xi_f,\chi,\theta_B')$.

In the top panel of Fig.~\ref{fig:B-profile}, we schematically show the geometry of the magnetic field and how it varies with the anisotropy 
parameter $\xi$. In the bottom panel, we show the radial profile of the magnetic field and its anisotropy in the shocked region, as a function of 
the distance $(R-r)$ from the forward shock.

\begin{figure}
    \centering
    \includegraphics[width=0.473\textwidth]{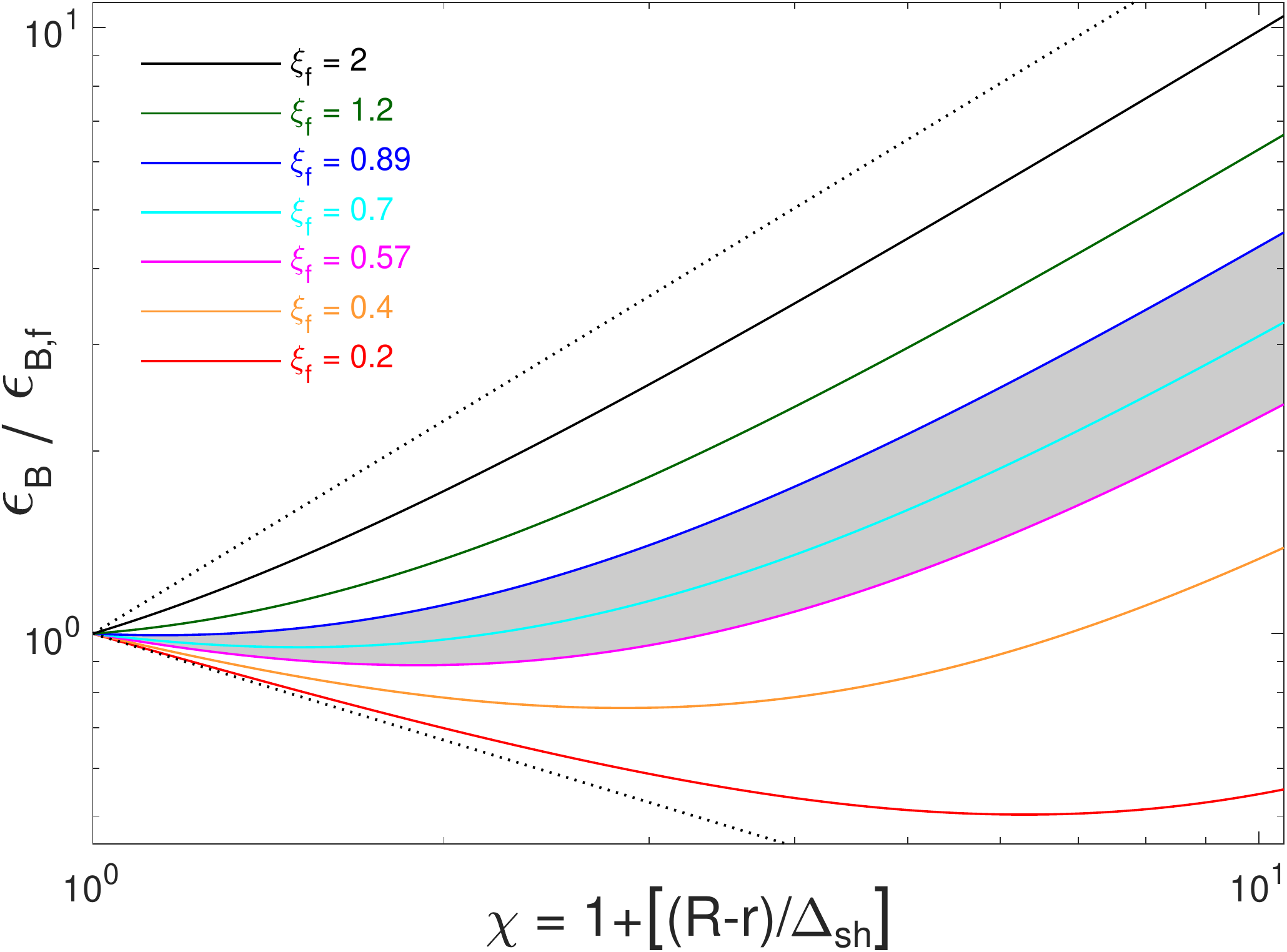}
    \caption{The evolution of the magnetic field equipartition parameter, $\epsilon_B$, with the distance behind the shock, according to Eq.~(\ref{eq:epsilon_B-flux_freezing}). The value of $\epsilon_B$ normalized by that just after the shock is shown as a function of the \citet[][]{Blandford-McKee-76} self-similar variable $\chi=1+[(R-r)/\Delta_{\rm sh}]$ where $\Delta_{\rm sh}=R/[2(4-k)\Gamma_{\rm sh}^2]$, for several different values of the magnetic field anisotropy parameter just behind the shock, $\xi_f=0,\,0.2,\,0.4,\,0.57,\,0.7,\,0.89,\,1.2,\,2,\,\infty$ (from bottom to top). The two extreme values of $\xi_f=0,\,\infty$ are shown as dotted (straight) lines and correspond to power laws in $\chi$, as given in Eq.~(\ref{eq:epsilon_B-flux_freezing}). The light-gray shaded region corresponds to the allowed range that we find in this work, $0.57\lesssim\xi_f\lesssim0.89$.}
    \label{fig:epsilon_B-chi}
\end{figure}

The radial structure of the two magnetic field components from Eq.~(\ref{eq:B-perp-par-chi}) can now be used, 
along with Eq.~(\ref{eq:B-chi} \& \ref{eq:xi}), to express the radial scaling of the magnetic field microphysical 
parameter in the downstream region (\citealt{Granot+99,Granot+99b}; also see appendix (\ref{sec:appendix}) for 
derivation),
\begin{equation}\label{eq:epsilon_B-flux_freezing}
 \frac{\epsilon_B}{\epsilon_{B,f}} = 
 \frac{2+\xi_f^2\chi^\frac{7-2k}{4-k}}{\left(2+\xi_f^2\right)\,\chi^{\frac{7-2k}{3(4-k)}}}\longrightarrow
 \left\{
    \begin{array}{cc}
        \chi^{-\frac{7-2k}{3(4-k)}}\ \ {\rm for} & \xi_f=0\ \  (B=B_\perp)\ , \\
        \chi^{\frac{2(7-2k)}{3(4-k)}}\ \ {\rm for} & \xi_f=\infty\ \  (B=B_\parallel)\ . 
    \end{array}\right.
\end{equation}
Figure~\ref{fig:epsilon_B-chi} shows $\epsilon_B(\chi)/\epsilon_{B,f}$ for several values of $\xi_f$. For $\xi_f>1$, 
$\epsilon_B$ monotonically increases with $\chi$ (and hence with the distance behind the shock), while for $\xi_f<1$, 
$\epsilon_B$ first decreases with $\chi$ until reaching $\xi=1$ and then increases with $\chi$.

\section{Linear Polarization}
\label{sec:lin-pol}

To calculate the linear polarization averaged over the entire afterglow image on the plane of the sky, we start 
by first expressing the flux density measured by an off-axis observer whose LOS points in the 
direction of the unit vector $\hat{\mathbfit{n}}'$ that makes an angle $\theta_{\rm obs}$ with the jet symmetry axis.

Here we consider synchrotron emission produced by relativistic electrons (or $e^\pm$-pairs) that are accelerated 
at the forward shock into a power law energy distribution, with $dN/d\gamma\propto\gamma^{-p}$ for $\gamma_m\leq\gamma\leq\gamma_M$. 
The comoving synchrotron emission coefficient (emitted energy per unit volume, time, frequency and solid angle) 
from a point-like region is given by
\begin{equation}
 j'_{\nu'}=j'_{\nu',0}\nu'^{-\alpha}\langle[B(\theta_B')\sin\delta']^\epsilon\rangle~,
\end{equation}
where it depends on the 
spectral index $\alpha$ and the component of the magnetic field perpendicular to the observer's LOS, $B\sin\delta'$, 
raised to some power $\epsilon=1+\alpha$ \citep{Laing-80}. Here $\delta'$ is the angle between the local comoving magnetic field 
unit vector $\hat{\mathbfit{B}}$ and the direction to the observer, $\hat{\mathbfit{n}}'\cdot\hat{\mathbfit{B}}=\cos\delta'$, 
and it depends on the angle between the direction to the observer and the shock normal, 
given by $\tilde\theta'=\arccos(\hat{\mathbfit{n}}'\cdot\hat{\mathbfit{n}}_{\rm sh})$. 
The normalization, $j'_{\nu',0} = j'_{\nu',0}(\theta,y,R)$, depends on the energy density and number density of the 
radiating electrons \citep[see, e.g.,][]{Gill-Granot-18b}. 
Since the magnetic field is tangled up in 3D with some anisotropy, the emissivity from a point-like region must be obtained by 
averaging over the different directions of the magnetic field, as indicated by the $\langle\rangle$ brackets, such that
\begin{equation}\label{eq:jnu-ave}
    \langle [B(\theta_B')\sin\delta']^\epsilon\rangle = \frac{\int[B(\theta_B')\sin\delta']^\epsilon f(\theta_B')d\Omega_B'}
    {\int f(\theta_B')d\Omega_B'}~,
\end{equation}
where the solid angle $d\Omega_B' = \sin\theta_B'd\theta_B'd\varphi_B'$ and $\varphi_B'$ is the azimuthal angle. Since the magnetic 
field distribution is symmetric around $\hat{\mathbfit{n}}_{\rm sh}$, this average would depend only on the angle 
$\tilde\theta'$ in addition to $\epsilon$ and $\xi$ (and these dependencies carry through to $j'_{\nu'}$). The flux 
density measured by an off-axis observer from a distant source at redshift $z$, corresponding to a luminosity distance 
$d_L$, is given by integrating over the volume of the emission region \citep[e.g.][]{Granot+99},
\begin{equation}
    F_\nu(t_{\rm obs}, \theta_{\rm obs}, \nu) = \frac{(1+z)}{d_L^2}\int_0^{2\pi}d\tilde\varphi
    \int_{-1}^1d\tilde\mu\int_0^1dy\left\vert\frac{dr}{dy}\right\vert\frac{(yR)^2j'_{\nu'}}{[\Gamma(1-\beta\tilde\mu)]^2}\,,
\end{equation}
where $\Gamma$ is the LF of the radiating fluid element, 
$\tilde{\mu}=\cos\tilde{\theta}=\hat{\mathbfit{n}}\cdot\hat{\boldsymbol{\beta}}=\hat{\mathbfit{n}}\cdot\hat{\mathbfit{r}}$, and $\tilde\theta$ and $\tilde\varphi$ 
are, respectively, the polar angle measured from and the azimuthal angle measured around the observer's LOS.

\begin{figure}
    \centering
    \includegraphics[width=0.477\textwidth]{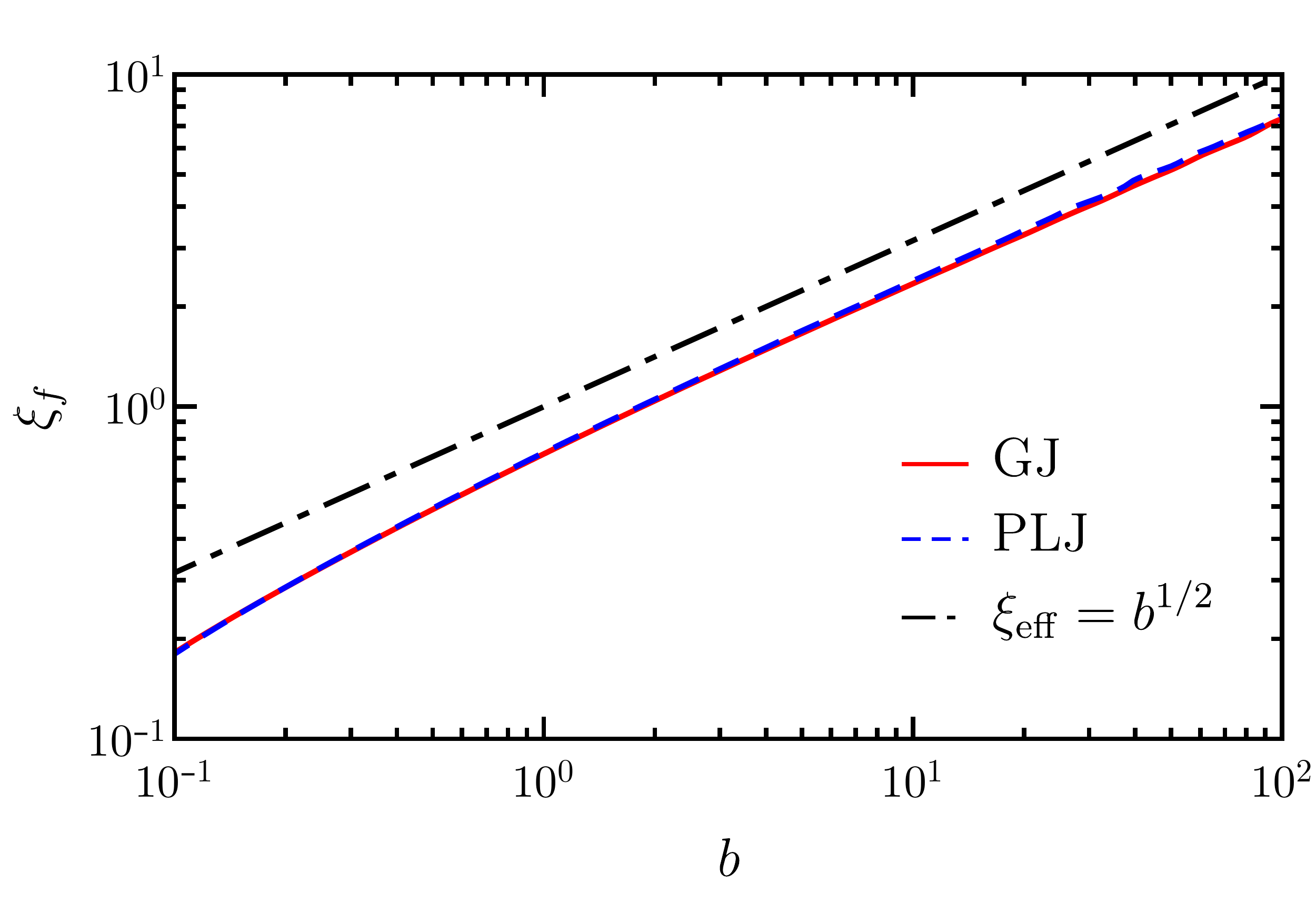}
    \caption{Mapping between the two parameters, $b$ and $\xi_f$, that characterize the 
    degree of anisotropy in the postshock magnetic field, shown at $t_{\rm obs}=244\,$days. 
    This mapping is not constant in time. The $b=2\langle B_\parallel^2\rangle/\langle B_\perp^2\rangle$ 
    parameter averages over both the radial profile and the local direction and strength distribution 
    of the field and is used in 2D integrals of the emission region, which is assumed to be an infinitely 
    thin shell. The $\xi_f$ parameter characterizes the anisotropy of the magnetic field just behind 
    the shock and is used in 3D volume integrals of the emission region. The scaling of the local effective 
    field anisotropy ($\xi_{\rm eff}$) with $b$ is also shown (also see the dotted lines in Fig~\ref{fig:xif-pol-244days}).}
    \label{fig:b-xif}
\end{figure}

The measured linear polarization ($V=0$) is obtained from the ratio of the polarized intensity to the total intensity, 
$\Pi = \sqrt{Q^2+U^2}/I$, where $I,Q,U,V$ are the Stokes parameters. Due to the 
assumed axisymmetry of the flow, $U$ vanishes, and the frequency independent polarization is therefore given by
\begin{equation}
    \Pi(t_{\rm obs},\theta_{\rm obs}) = \frac{Q}{I} = \frac{\int dF_\nu\cos(2\tilde\varphi)\Pi'}{\int dF_\nu}~,
\end{equation}
with $\Pi'$ being the local polarization. 
When $\Pi < 0$ ($\Pi > 0$) the polarization vector on the plane of the sky is expected to be along (normal to) the 
direction of motion of the flux centroid. To obtain the local polarization $\Pi'$ from a point-like region averaging 
over the different magnetic field directions is again performed, which yields \citep{Sari-99}
\begin{equation}\label{eq:local-pol}
    \frac{\Pi'}{\Pi_{\max}} = \frac{\int \cos(2\eta')[B(\theta_B')\sin\delta']^\epsilon f(\theta_B')d\Omega_B'}
    {\int [B(\theta_B')\sin\delta']^\epsilon f(\theta_B')d\Omega_B'}~,
\end{equation}
where $\eta'$ is the angle between the direction of the local polarization unit vector 
$\hat{\boldsymbol{\Pi}}'=(\hat{\mathbfit{n}}'\times\hat{\mathbfit{B}})/\vert\hat{\mathbfit{n}}'\times\hat{\mathbfit{B}}\vert$ 
and the direction perpendicular to the plane containing $\hat{\mathbfit{n}}'$ and $\hat{\boldsymbol{\beta}}$. 
When $\Pi'>0$ ($\Pi'<0$) the direction of the local polarization vector is along (normal to) the direction of 
$\hat{\mathbfit{n}}\times\hat{\mathbfit{n}}_{\rm sh}$. Again, because of the magnetic field's symmetry around 
$\hat{\mathbfit{n}}_{\rm sh}$, this ratio depends only on $\tilde\theta'$, $\epsilon$ and $\xi$. For $\epsilon=2$ (with $\alpha=1$), 
the  expression for the polarization becomes particularly simple,
and Eq.~(\ref{eq:local-pol}) yields \citep{Gruzinov-99,Sari-99}
\begin{equation}
    \frac{\Pi'(\tilde\theta')}{\Pi_{\max}} = \frac{(b-1)\sin^2\tilde\theta'}{2+(b-1)\sin^2\tilde\theta'}
    \quad\quad\quad(\epsilon=2)~,
\end{equation}
where $\cos\tilde\theta' \equiv \tilde\mu' = (\tilde\mu-\beta)/(1-\beta\tilde\mu)$. Here 
$b\equiv2\langle B_\parallel^2\rangle/\langle B_\perp^2\rangle$ is another way to parameterize the anisotropy of the magnetic field
\citep{Granot-Konigl-03}, where the average is taken over the radial profile of the flow downstream of the shock and the local 
direction and strength distribution of the magnetic field. 
This choice of parameterization is most useful when considering emission from an infinitely thin region behind the shock 
\citep[see, e.g.,][]{Granot-Konigl-03}, for which the degree of polarization is obtained from
\begin{equation}
    \Pi_{\rm 2D}=\frac{Q}{I} = \frac{\int\delta_D^3L'_{\nu'}\langle [B(\theta_B')\sin\delta']^2\rangle\Pi'\cos(2\tilde\varphi)d\tilde\Omega}
    {\int\delta_D^3L'_{\nu'}\langle [B(\theta_B')\sin\delta']^2\rangle d\tilde\Omega}\quad(\epsilon=2)~,
\end{equation}
where $d\tilde\Omega=d\tilde\mu d\tilde\varphi$ and $L'_{\nu'}$ is the isotropic equivalent (locally anisotropic) synchrotron spectral power 
\citep[see][for further details]{Gill-Granot-18b}. Averaging over different magnetic field directions follows from 
Eq.~(\ref{eq:jnu-ave}), which yields\footnote{This factor was neglected in \citet{Gill-Granot-18b} and 
only an isotropic comoving synchrotron spectral luminosity was assumed there.}
\begin{equation}
    \langle [B(\theta_B')\sin\delta']^2\rangle \propto 2+(b-1)\sin^2\tilde\theta'\quad\quad\quad (\epsilon=2)~.
\end{equation}
In Fig.~\ref{fig:b-xif} we show the mapping between $b$ and $\xi_f$ that was obtained by comparing the degree of 
polarization in the two cases for the two different jet structures at $t_{\rm obs}=244\,$days. Note that applying 
the local definition of $b$ to our local magnetic field model that is defined by $\xi$ alone yields the simple 
relation $b\equiv2\langle B_\parallel^2\rangle/\langle B_\perp^2\rangle=\xi^2$ (see appendix~(\ref{sec:appendix})). 
This local relation largely carries 
through to the global mapping between $b$ and $\xi_f$ shown in Fig.~\ref{fig:b-xif}, when accounting for the fact 
that since $\xi$ increases with the distance behind the shock, its effective value that can be defined as 
$\xi_{\rm eff}=b^{1/2}$ is somewhat higher than that just behind the shock ($\xi_f$).

\begin{figure}
    \centering
    \includegraphics[width=0.477\textwidth]{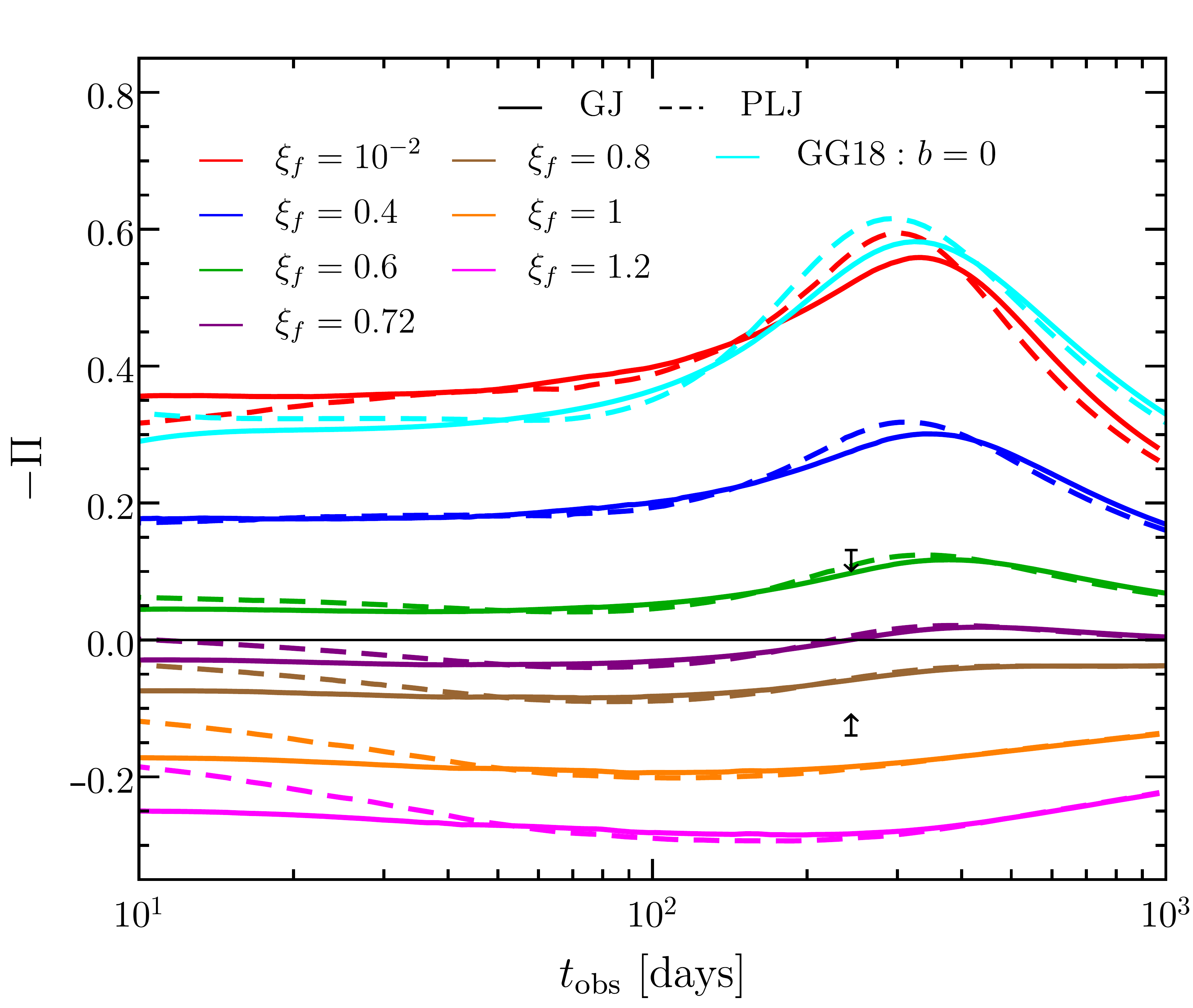}
    \caption{Temporal evolution of the degree of linear polarization ($\Pi$), obtained from a volume integration of the 
    flow, shown for different values of the magnetic field anisotropy parameter, $\xi_f$, just behind the shock. The two arrows 
    mark the polarization upper limit, $\vert\Pi\vert < 12\%$. Comparison 
    is made between two jet structures -- a gaussian jet (GJ) and a power law jet (PLJ). The result from \citet[][GG18]{Gill-Granot-18b}, 
    which assumed an infinitely thin shell geometry as well as locally isotropic synchrotron spectral emissivity (see footnote 2), is also shown 
    for the magnetic field anisotropy parameter $b=0$. The jet and afterglow model parameters are the same as assumed in GG18.}
    \label{fig:xif-pol}
\end{figure}

We present the temporal evolution of the linear polarization for different $\xi_f$ parameter values and the two jet 
structures in Fig.~\ref{fig:xif-pol}. The polarization curve obtained from the 2D surface integral for 
$b=0$ in GG18 is also shown for comparison. As shown in GG18, the peak of the polarization occurs at $\sim2t_{\rm obs,pk}$, 
where $t_{\rm obs,pk}\approx150\,$days is the peak of the lightcurve when the compact relativistic core of the outflow becomes 
visible to the off-axis observer. This remains true for $\xi_f<1$. As the two parameters are 
increased above zero the degree of anisotropy of the magnetic field begins to decline, until the field becomes completely isotropic, 
which occurs at $b=1$ and $\xi_f\approx0.72$ (corresponding to $\xi_{\rm eff}=1$) in the two cases. This also marks the point 
when the polarization vanishes and the polarization position angle 
undergoes a $90^\circ$ flip. Prior to this point, as the magnetic field anisotropy decreases the level of polarization also declines. 
The trend reverses post this point, when the $B_\parallel$ components starts to dominate over the $B_\perp$ component and as the 
field again becomes increasingly anisotropic.

An upper limit on the degree of linear polarization of $\vert\Pi\vert<12\%$ ($99\%$ confidence) was measured by \citet{Corsi+18} 
from the radio afterglow of \gw\ using the VLA at $t_{\rm obs}\approx244\,$days and $\nu=2.8\;$GHz.\footnote{Detailed modeling of 
the  GRB$\,$170817A/GW$\,$170817 afterglow shows that the observed frequency is well within PLS G so it is not affected by 
synchrotron self-absorption. This suggests that plasma propagation effects in the source are also negligible at the time of this observation.} Since the polarization angle could not be constrained, the upper 
limit, shown as two black arrows in Fig.~\ref{fig:xif-pol}, constrains the absolute value of the true degree of polarization. We use 
this upper limit to constrain the value of both $0.66\lesssim b\lesssim1.49$ and $0.57\lesssim\xi_f\lesssim0.89$ in Fig.~\ref{fig:xif-pol-244days}, 
where we show the degree of polarization at $t_{\rm obs}=244\,$days as a function of $b$ and $\xi_f$ for the two jet structures. The postshock 
field anisotropy ($\xi_f$) may have some dependence on the shock compression ratio, $\rho_f/\rho_k(r)=2^{3/2}\Gamma_{\rm sh}=4\Gamma_f$ 
(see Eq.~(\ref{eq:rho-chi}); where the last equality is true when $\Gamma_{\rm sh}\gg1$). The corresponding ranges for the flux-weighted mean 
of the two LFs and for the two jet structures are: (GJ) $4.26\leq\langle\Gamma_{\rm sh}\rangle\leq4.59$; $3.31\leq\langle\Gamma_f\rangle\leq3.59$, 
(PLJ) $3.98\leq\langle\Gamma_{\rm sh}\rangle\leq4.24$; $3.10\leq\langle\Gamma_f\rangle\leq3.30$. The dotted lines in Fig.~\ref{fig:xif-pol} show 
$\Pi(\xi_f^2)$ that have the same trend as $\Pi(b)$. This similarity results from the local scaling (averaged over the field's angular distribution) 
where $b=\xi^2$. It is preserved in the global 3D integration where the effective anisotropy of the postshock field scales as $\xi_{\rm eff}=b^{1/2}$.

\section{Discussion}
\label{sec:discussion}
\begin{figure}
    \centering
    \includegraphics[width=0.477\textwidth]{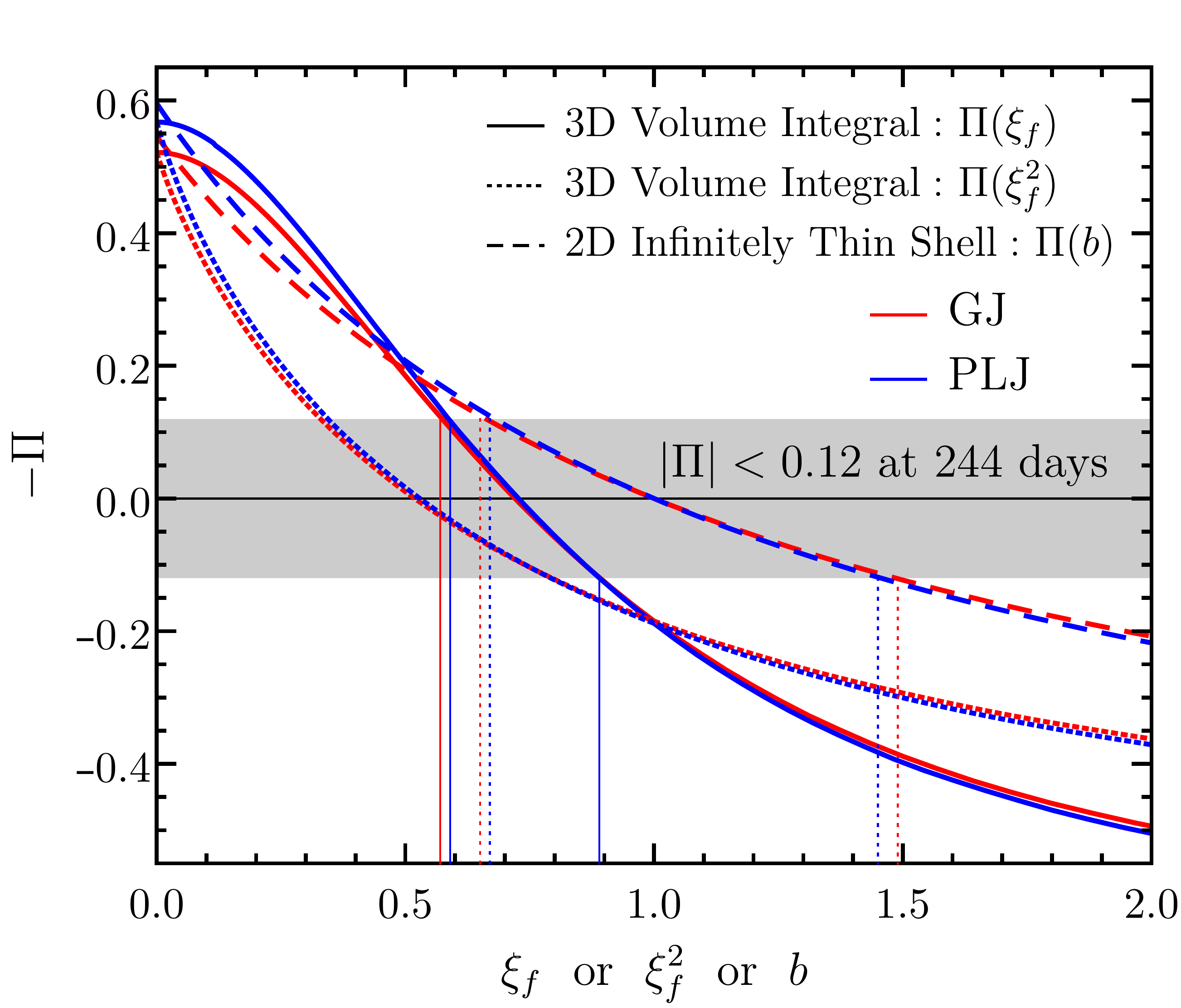}
    \caption{Linear polarization ($\Pi$) as a function of the magnetic field anisotropy parameter just behind the shock $\xi_f$ for a 3D volume integral or the anisotropy paremeter $b$ for a 2D infinitely thin shell \citep{Gill-Granot-18b}. 
    The upper limit on $\vert\Pi\vert$ measured by \citet{Corsi+18} at $t_{\rm obs}\approx244\,$days post merger is shown 
    as the gray shaded region. 
    This constrains $0.57\lesssim\xi_f\lesssim0.89$ and $0.66\lesssim b\lesssim 1.49$ for both 
    the Gaussian jet (GJ; $0.57\lesssim\xi_f\lesssim0.89$, $0.66\lesssim b\lesssim 1.49$) and power law 
    jet (PLJ; $0.59\lesssim\xi_f\lesssim0.89$, $0.68\lesssim b\lesssim 1.46$). The dotted lines show $\Pi(\xi_f^2)$ having the 
    same shape as $\Pi(b)$ demonstrating that, since locally $b=\xi^2$, a global 3D integration preserves the same scaling for 
    an effective anisotropy parameter, $\xi_{\rm eff}=b^{1/2}$.}
    \label{fig:xif-pol-244days}
\end{figure}

The upper limit of $\vert\Pi\vert<12\%$ measured for the afterglow linear polarization for \gw\ has important 
implications for the postshock magnetic field structure, in particular for its degree of anisotropy. In the case of Weibel 
instability-generated magnetic fields just behind the forward shock, the theoretically expected value \citep[e.g.,][]{Medvedev-Loeb-99} of the 
anisotropy parameter is $\xi_f=0$ (corresponding to $b=0$ in the 2D case). As shown in Fig.~\ref{fig:xif-pol-244days}, this case is 
ruled out and the constrained value of $0.57\lesssim\xi_f\lesssim0.89$ suggests that the magnetic field just behind the shock must have 
a finite but sub-dominant $B_\parallel$ component. In terms of the parameter $b$, which is more suitable for a 2D thin emitting shell, we obtain $0.66\lesssim b\lesssim 1.49$. This is a significant improvement compared to the previous rough estimate 
of $0.5\lesssim b\lesssim2$ made by \citet{Granot-Konigl-03} based on the 
low measured values ($\Pi\lesssim\;$few \% in the optical/NIR) of afterglow linear polarization in the first several GRB afterglows (which were available at that time), which involved a statistical argument since the jet angular structure and $\theta_{\rm obs}$ were not known for those GRBs.

On the other hand, 3D PIC simulations of two counterstreaming unmagnetized relativistic electron-ion (or even electron-positron) 
plasmas find that the magnetic field just behind the shock is predominantly transverse, with a finite component parallel to the 
shock normal having, on average, $\vert B_\parallel\vert/\vert B\vert\lesssim10^{-2}$ \citep[e.g.,][]{Frederiksen+04,Kazem+15,Kazem+16}. 
Expressing this ratio in terms of 
the anisotropy parameter, we find $\vert B_{\parallel,f}\vert/\vert B_f\vert = \xi_f(1+\xi_f^2)^{-1/2} \simeq 0.50 - 0.66$ for 
$0.57\lesssim\xi_f\lesssim0.89$, which suggests that the field anisotropy just behind the shock is significantly smaller as compared 
to that found in those PIC simulations. Furthermore, due to the larger stretching of the flow in the radial direction (compared to 
the two transverse directions) $\vert B_{\parallel,f}\vert/\vert B_f\vert$ grows with the distance behind the shock (however this 
occurs on the dynamical scales where the planar symmetry of the PIC simulations breaks down and the shock's radius of curvature becomes 
important).
Since PIC simulations are generally limited to box-sizes that span at most 
$\lesssim10^4(c/\omega_{p,e})$ \citep[see, e.g.,][for 2D simulations]{Spitkovsky-08a,Keshet+09}, which is still much smaller than the width of the 
postshock region, the decline in field anisotropy over larger scales remains unconstrained. 

In modeling the afterglow, we have explicitly made the assumption that $\epsilon_B$ (with $\epsilon_{B,f}=5.5\times10^{-4}$ in this work, 
though it is still subject to degeneracies with other model parameters, e.g. \citealt{Gill+19}) depends on the radial profile of the 
flow in the shocked region. PIC simulations show that the two-stream/Weibel instability amplifies the magnitude of small scale shock-generated magnetic 
field to near-equipartition ($\epsilon_B\sim0.1$) in the shock transition region that separates the upstream and downstream flows. However, 
many 2D PIC simulations find that due to particle phase-space mixing the field decays rapidly downstream within few $\times\,100(c/\omega_{p,e})$ 
\citep[e.g.,][]{Kato-07,Chang+08,Spitkovsky-08a}. The long-term evolution of both electron-ion \citep{Takamoto+18} and $e^\pm$-pair plasma 
\citep{Keshet+09} PIC simulations does seem to suggest that the magnetic field decay saturates at $\epsilon_B\sim10^{-2}$ for comoving distances 
$\gtrsim10^2(c/\omega_{p,e})$ from the shock transition region. However, due to the small dynamical scales probed in these simulations and the 
assumption of planar symmetry, the radial stretching of fluid elements is not observed. Therefore, a self-consistent treatment assuming flux freezing 
once plasma effects saturate and $\xi_f$ with the corresponding $\epsilon_{B,f}$ is established, would require $\epsilon_B$ to evolve with 
$\chi$ according to Eq.~(\ref{eq:epsilon_B-flux_freezing}).  

Early 2D PIC simulations that showed a gradual power law decay of $\epsilon_B$ prompted the consideration of decaying magnetic fields in 
the bulk of the shocked region for afterglow modeling \citep{Rossi-Rees-03,Lemoine+13,Lemoine-15} and also in models of prompt emission from internal shocks 
\citep{Peer-Zhang-06,Derishev-07}. Moreover, afterglow modeling of many GRBs in the optical band revealed a wide distribution with a rather 
low value of the radially averaged magnetic field in the shocked region with $10^{-8}\lesssim\langle\epsilon_B\rangle\lesssim10^{-2}$ under the 
explicit assumption that the circumburst density $n=1\,{\rm cm}^{-3}$ \citep{Santana+14}. Therefore, even though long-term PIC simulations might 
hint at saturation of the field, as mentioned above, more realistic simulations for the GRB afterglow shock should show its evolution with $\xi$ 
in the downstream. 

Instead of generating magnetic field in the shock upstream via microscopic plasma instabilities (e.g., two-stream/Weibel), it has been argued 
\citep{Sironi-Goodman-07,Couch+08,Goodman-MacFadyen-08} and demonstrated using MHD simulations \citep{Zhang+09,Inoue+11,Mizuno+11} that macroscopic 
turbulence can amplify  the feeble ($\epsilon_B\sim10^{-9}$) ISM magnetic field to $\epsilon_B\sim\Gamma^{-1}\gtrsim10^{-2}$, with $\Gamma\lesssim{\rm few}\times10^2$ for GRB afterglows, via vorticity generation at the shock transition due to density inhomogeneities in the shock upstream. 
As the cold density clumps pass through the shock transition, vortical eddies 
are created in the downstream that twist and stretch the seed magnetic field. This leads to amplification of the field strength over the eddy turn-over 
time due to the dynamo mechanism. The density clumps in the upstream can arise either from preexisting density inhomogeneities in the stellar wind 
(relevant only to long-soft GRBs) or can be self-generated due to partial charge separation between the shock accelerated non-thermal ions and electrons 
in the upstream \citep{Couch+08}. The advantage of turbulence amplified magnetic field is that its coherence length is much larger than plasma skin-depth 
scales, making it less susceptible to decay due to particle phase space mixing. Moreover, the postshock field tends to be more isotropic, which is 
consistent with the findings of this work.

\section{Conclusions}
\label{sec:conclusion}

In this work, we have used the upper limit on the linear polarization, $\vert\Pi\vert<12\%$ at $t_{\rm obs}\approx244\,$ days, 
of the radio afterglow of \gw\ to constrain the anisotropy of the shock-generated tangled magnetic field. 
The structure of the outflow was modeled using the best-fit solution (from afterglow data up to $t_{\rm obs}\sim600\,$days) 
obtained in \citet{Gill-Granot-18b,Gill+19} for a Gaussian and a power-law jet with locally spherical dynamics and no sideways 
spreading. Since the flux at the time of polarization measurement was dominated 
by the relativistic core, with $\Gamma(t_{\rm obs}=244\,{\rm days})\approx4.1(t_{\rm obs}/150\,{\rm days})^{-3/8}=3.4$, the assumption 
of no sideways spreading, which is important when the flow becomes non-relativistic \citep[e.g.,][]{vanEerten-MacFadyen-12b}, 
should still remain reasonably valid. As suggested by the broadband (X-ray/Optical/Radio) afterglow 
synchrotron emission, which maintained a single power law with $\nu_m<\nu<\nu_c$, the shock-heated relativistic power-law electrons were 
slow cooling. This necessitated the need for integrating over the entire shocked volume, rather than assuming an infinitely thin 
emission region behind the shock, to calculate the observed linear polarization. In order to do that, we have modeled the radial 
profile of the postshock magnetic field using the BM76 relativistic spherical self-similar solution. Under the frozen-field 
approximation, this causes the magnetic field to become increasingly radial since the farther downstream the flow is the more radially 
(as compared to the transverse direction) stretched a given fluid element becomes. 

Our main conclusion is that the shock-generated tangled magnetic field cannot lie only in the plane of the shock 
(perpendicular to the shock normal, $B_\perp$), as posited by some theoretical works \citep[e.g.,][]{Medvedev-Loeb-99} 
and also shown in some PIC simulations \citep[e.g.,][]{Chang+08}. We find that the field just behind the shock must have a finite, 
albeit mildly sub-dominant, component parallel to the shock normal, $B_\parallel$, which is radial in our case. Moreover, 
the initial field anisotropy parameter must be in the range $0.57\lesssim(\xi_f=B_{\parallel,f}/B_{\perp,f})\lesssim0.89$, 
and $\xi=B_{\parallel}/B_{\perp}=\xi_f\chi^{(7-2k)/(8-2k)}$ grows downstream with the distance behind the shock. This presents 
a mismatch between our results and that obtained both from analytic arguments and current PIC simulations of relativistic 
collisionless shocks, suggesting that larger scale and long-term simulations are needed to better constrain the asymptotic 
structure of the postshock magnetic field. At the same time, the lower degree of polarization of afterglow emission that reflects 
the inherent higher level of isotropy of the postshock magnetic field may be consistent with turbulence amplified magnetic field.

\section*{Acknowledgements}
This research was supported by the Israeli Science Foundation (ISF grant No. 719/14) and by the ISF-NSFC joint research program (grant No. 3296/19). We thank Ehud Nakar and Uri Keshet for useful discussions.



\appendix

\section{Magnetic field evolution as it is advected downstream of the shock}
\label{sec:appendix}

Here we derive the evolution of the magnetic field as it is advected downstream of the shock, assuming a local (i.e. at each given angle $\theta$ from the jet symmetry axis) \citet[][]{Blandford-McKee-76} radial dynamics.
Let a subscript ``0'' denote the value of a quantity when the fluid element that is currently (at a local shock radius $R$) at $\chi=1+2(4-k)\Gamma_{\rm sh}^2(R)[(R-r)/R]$, just crossed the shock (and was at $\chi=1$ when the shock radius was $R_0$), while a subscript ``$f$'' denotes the current (at a shock radius $R$) value of a quantity just behind the shock (at $\chi=1$). We have \citep{Granot+99,Granot-Sari-02} $\chi=(R/R_0)^{4-k}$
while the proper internal energy density scales as
\begin{equation}\label{eq:e_evolution}
    \frac{e}{e_0} = \chi^{-\frac{2(13-2k)}{3(4-k)}}\ ,\quad\quad
    \frac{e_f}{e_0} = \chi^{-\frac{3}{4-k}}\ .
\end{equation}
The comoving length of a fluid element in the direction parallel to the shock normal $\hat{\mathbfit{n}}_{\rm sh}$ and the two directions perpendicular to it scale as \citep{Granot+99,Granot-Konigl-03}
\begin{equation}
    \frac{L_\parallel}{L_{\parallel,0}}=\chi^{\frac{9-2k}{2(4-k)}}\ ,\quad\quad
    \frac{L_\perp}{L_{\perp,0}}=\frac{R}{R_0}=\chi^{\frac{1}{4-k}}\ .
\end{equation}
Assuming flux freezing as the fluid element is advected downstream, the two corresponding components of the comoving magnetic field scale as
\begin{equation}\label{eq:B_scaling}
    \frac{B_\parallel}{B_{\parallel,0}}=\frac{L_{\perp,0}^2}{L_\perp^2}=\chi^{-\frac{2}{4-k}}\ ,\quad\quad
    \frac{B_\perp}{B_{\perp,0}}=\frac{L_{\perp,0}L_{\parallel,0}}{L_{\perp}L_{\parallel}}=\chi^{-\frac{11-2k}{2(4-k)}}\ .
\end{equation}

Now, in order to calculate the evolution of $\epsilon_B = \mean{B^2}/8\pi e$
we need to calculate the mean of the square of the magnetic field. In our formalism the magnetic field is derived from an isotropic 
distribution that we will denote by a bar, with constant magnetic field strength $\bar{B}$ and angular probability density 
$\bar{f}=1/4\pi$ so that $\int\bar{f}d\bar{\Omega}'_B=1$ where $d\bar{\Omega}'_B=d\bar{\varphi}'_B d\bar{\mu}$ and $\bar{\mu}=\cos\bar{\theta}'_B$. 
Since the field is symmetric with respect to $\hat{\mathbfit{n}}_{\rm sh}$, i.e. $\bar{f}=f(\hat{\theta}'_B)$, then we can integrate over $\bar{\varphi}'_B$ 
and switch from $\bar{f}d\bar{\Omega}'_B$ to $\bar{f}_\mu d\bar{\mu}$ where $\bar{f}_\mu=\frac{1}{2}$ is normalized such that 
$\int_{-1}^{1}\bar{f}_\mu d\bar{\mu}=1$. The magnetic field is derived from this distribution by stretching the component parallel 
to $\hat{\mathbfit{n}}_{\rm sh}$ by a factor $\xi$ while the perpendicular component remains unchanged,
\begin{equation}
    \xi=\frac{B_\parallel}{\bar{B}_\parallel}=\frac{B\mu}{\bar{B}\bar{\mu}}\ ,\quad\quad
    1=\frac{B_\perp}{\bar{B}_\perp}=\frac{B}{\bar{B}}\sqrt{\frac{1-\mu^2}{1-\bar{\mu}^2}}\ .
\end{equation}
This implies the relation
\begin{equation}
    \bar{\mu}=\left[1+\xi^2(\mu^{-2}-1)\right]^{-1/2}\ ,
\end{equation}
and therefore the post-stretching magnetic field strength as a function of $\mu=\cos\theta'_B=\hat{\mathbfit{n}}_{\rm sh}\cdot\hat{\mathbfit{B}}$ is
\begin{equation}
    \frac{B}{\bar{B}}=\sqrt{\frac{B_\parallel^2+B_\perp^2}{\bar{B}^2}}=
    \frac{\xi}{\sqrt{\xi^2(1-\mu^2)+\mu^2}}\ .
\end{equation}
The implied post-stretching angular probability distribution is
\begin{equation}
    f_\mu(\mu)=\bar{f}_\mu\frac{d\bar{\mu}}{d\mu}
    =\frac{\xi^2}{2}\fracb{\bar{\mu}}{\mu}^3
    =\frac{1}{2\xi}\fracb{B}{\bar{B}}^3
    =\frac{\frac{1}{2}\xi^2}{\left[\xi^2(1-\mu^2)+\mu^2\right]^{3/2}}\ .
\end{equation}
Now it is straightforward to calculate the mean of any quantity $Q$ over the magnetic field as $\mean{Q}=\int_{-1}^{1}d\mu f_\mu(\mu)\,Q$. 
Note that more generally, when $Q$ also depends on $\varphi=\varphi'_B$, one needs to evaluate 
$\mean{Q}=\int_{0}^{2\pi}d\varphi\int_{-1}^{1}d\mu f(\mu)\,Q(\mu,\varphi)$ where $f(\mu)=f_\mu(\mu)/2\pi$. In particular, the mean 
of the square of the parallel and perpendicular field components, normalized by their post-stretching values, are
\begin{equation}\label{eq:B_square_mean}
\frac{\mean{B_\parallel^2}}{\mean{\bar{B}^2}}=\frac{\xi^2}{3}\ ,\quad\quad
\frac{\mean{B_\perp^2}}{\mean{\bar{B}^2}}=\frac{2}{3}\ ,\quad\quad
\frac{\mean{B^2}}{\mean{\bar{B}^2}}=\frac{\mean{B_\perp^2}+\mean{B_\parallel^2}}{\mean{\bar{B}^2}}=\frac{2+\xi^2}{3}\ .
\end{equation}
This result can be understood in a simple way. Before the stretching the field is isotropic so each of the three directions holds $\frac{1}{3}$ of $\mean{\bar{B}^2}=\bar{B}^2$, and since there are two perpendicular directions and one parallel direction then $\mean{\bar{B}_\perp^2}=\frac{2}{3}\mean{\bar{B}^2}$ and $\mean{\bar{B}_\parallel^2}=\frac{1}{3}\mean{\bar{B}^2}$. Since the perpendicular component remains unchanged ($B_\perp=\bar{B}_\perp$) so does the mean of its square, and since the parallel component changes by a factor of $\xi$ then its square changes by a factor of $\xi^2$ everywhere, and so does its mean value.
This result immediately gives us the local value of the parameter $b$,
\begin{equation}
    b\equiv\frac{2\mean{B_\parallel^2}}{\mean{B_\perp^2}}=\xi^2\ ,\quad\quad
    b(\,\chi) = \xi^2(\,\chi) = \xi_f^2\,\chi^{\frac{7-2k}{4-k}}\ .
\end{equation}

In our formalism, the local pre-stretching magnetic field strength corresponds to $\bar{B}\to B_{\perp,\rm{max}}=B(\mu=0)$. According to Eq.~(\ref{eq:B_scaling}), $B_{\perp,\rm{max}}^2/B_{\perp,\rm{max},0}^2=\chi^{-(11-2k)/(4-k)}$, and therefore
\begin{equation}\label{eq:B_square_evolution}
    \frac{\mean{B^2}}{\mean{B_0^2}}=\frac{2+\xi^2}{2+\xi_0^2}\frac{B_{\perp,\rm{max}}^2}{B_{\perp,\rm{max},0}^2}=\frac{2+\xi_f^2\,\chi^{\frac{7-2k}{4-k}}}{2+\xi_f^2}\,\chi^{-\frac{11-2k}{4-k}}\ .
\end{equation}
We assume here for simplicity that both $\xi$ and $\epsilon_B$ have a universal value just behind the shock (at $\chi=1$), so that $\xi_0\equiv\xi(R_0,\chi=1)=\xi(R,\chi=1)\equiv\xi_f$ and $\epsilon_{B,0}\equiv\epsilon_B(R_0,\chi=1)=\epsilon_B(R,\chi=1)\equiv\epsilon_{B,f}$. This in not obvious since $\epsilon_{B,f}$ may vary with $\Gamma_{\rm sh}$ in the mildly relativistic regime, while $\xi_f$ may even vary in the ultra-relativistic regime, if e.g. it depends on the shock compression ratio.
Anyway, under our assumptions, using Eqs.~(\ref{eq:e_evolution}) and (\ref{eq:B_square_evolution}) the evolution of $\epsilon_B$ is given by
\begin{equation}
    \frac{\epsilon_B}{\epsilon_{B,f}}=\frac{\epsilon_B}{\epsilon_{B,0}}
    =\frac{\mean{B^2}}{\mean{B_0^2}}\frac{e_0}{e}
    =\frac{2+\xi_f^2\,\chi^\frac{7-2k}{4-k}}{\left(2+\xi_f^2\right)\,\chi^{\frac{7-2k}{3(4-k)}}}\ .
\end{equation}
Similarly, the magnetic field just behind the shock is given by $B_{\perp,\rm{max},f}^2/B_{\perp,\rm{max},0}^2=
    \epsilon_{B,f}e_f/\epsilon_{B,0}e_0=
    e_f/e_0=\chi^{-\frac{3}{4-k}}$, so that
\begin{equation}
    \frac{B_{\perp,\rm{max}}}{B_{\perp,\rm{max},f}}
    =\frac{B_{\perp,\rm{max}}}{B_{\perp,\rm{max},0}}\frac{B_{\perp,\rm{max},0}}{B_{\perp,\rm{max},f}}
    =\chi^{-1}\ ,
\end{equation}
\begin{equation}
    \frac{\mean{B_{\parallel}^2}}{\mean{B_f^2}}
    =\frac{\xi^2\chi^{-2}}{2+\xi_f^2}
    =\frac{\xi_f^2}{2+\xi_f^2}\,\chi^{-\frac{1}{4-k}}\ ,\quad\quad
    \frac{\mean{B_{\perp}^2}}{\mean{B_f^2}}
    =\frac{2\,\chi^{-2}}{2+\xi_f^2}\ .
\end{equation}
\begin{equation}
    \frac{\mean{B^2}}{\mean{B_f^2}}
    =\frac{2+\xi^2}{2+\xi_f^2}\,\chi^{-2}
    =\frac{2+\xi_f^2\,\chi^\frac{7-2k}{4-k}}{(2+\xi_f^2)\,\chi^2}\ .
\end{equation}

\bsp	
\label{lastpage}
\end{document}